\documentclass[reprint,prl,twocolumn,twoside,aps,superscriptaddress]{revtex4-2}

\usepackage{amsmath,amssymb,amsfonts}
\usepackage{graphicx}
\usepackage{siunitx}
\usepackage{natbib}
\usepackage{braket}
\usepackage{multirow}
\usepackage[normalem]{ulem}
\usepackage[dvipsnames]{xcolor}
\usepackage{xspace}
\usepackage{url}
\usepackage[breaklinks]{hyperref}
\usepackage{braket}

\hypersetup{
    unicode=true,
    colorlinks=true,
    linkcolor=blue,
    citecolor=blue,
    urlcolor=blue
  }
\usepackage{breakurl}
\usepackage{soul}

\newcommand{\Sstate}{$\ket{36\mathrm{S}_{1/2}, m_{j}=-1/2}_{\rm Rb}$\xspace}
\newcommand{\Dstate}{$\ket{56\mathrm{D}_{5/2}, m_{j}=5/2}_{\rm Rb}$\xspace}
\newcommand{\ground}{$\ket{g}_{\rm Rb}$\xspace}

\newcommand{\groundpair}{$\ket{g}_{\rm Rb}\ket{g}_{\rm Cs}$\xspace}
\newcommand{\intermediate}{$\ket{e}_{\rm Rb}$\xspace}
\newcommand{\ryd}{$\ket{r}_{\rm Rb}$\xspace}

\usepackage{xr}
\usepackage{pdfpages} 
\usepackage{pgffor} 

\makeatletter
\AtBeginDocument{\let\LS@rot\@undefined}
\makeatother

\def\supplementfilename{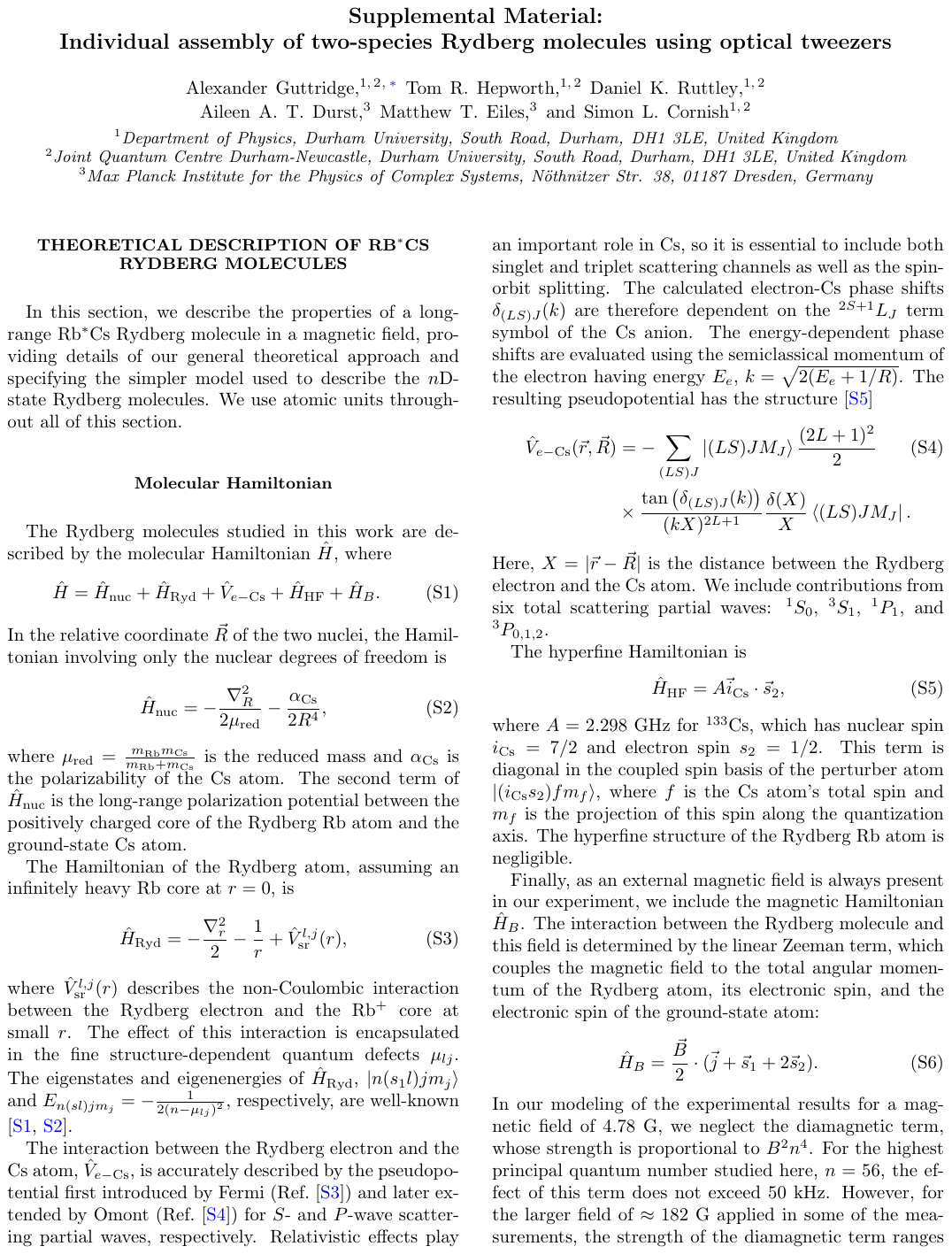}

\pdfximage{\supplementfilename}
\def\numbersupplementpages{\the\pdflastximagepages}

\newif\ifarXiv
\arXivtrue 

\begin{document}

\title{Individual assembly of two-species Rydberg molecules using optical tweezers}

\newcommand{\physics}{Department of Physics, Durham University, South Road, Durham, DH1 3LE, United Kingdom}
\newcommand{\jqc}{Joint Quantum Centre Durham-Newcastle, Durham University, South Road, Durham, DH1 3LE, United Kingdom}
\newcommand{\dresden}{Max Planck Institute for the Physics of Complex Systems, N\"othnitzer Str. 38, 01187 Dresden, Germany}

\author{Alexander Guttridge}
\email{alexander.guttridge@durham.ac.uk}
\affiliation{\physics}
\affiliation{\jqc}
\author{Tom R. Hepworth}
\affiliation{\physics}
\affiliation{\jqc}
\author{Daniel K. Ruttley}
\affiliation{\physics}
\affiliation{\jqc}
\author{Aileen A. T. Durst}
\affiliation{\dresden}
\author{Matthew T. Eiles}
\affiliation{\dresden}
\author{Simon L. Cornish}
\affiliation{\physics}
\affiliation{\jqc}

\begin{abstract}
We present a new approach to investigating Rydberg molecules by demonstrating the formation and characterization of individual Rb$^{*}$Cs Rydberg molecules using optical tweezers.
By employing single-atom detection of Rb and Cs, we observe molecule formation via correlated loss of both species and study the formation dynamics with single-particle resolution. 
We control the interatomic distances by manipulating the relative wavefunction of atom pairs using the tweezer intensity, optimizing the coupling to molecular states and exploring the effect of the tweezer on these states. Additionally, we demonstrate molecule association with atoms trapped in separate tweezers, paving the way for state-selective assembly of polyatomic molecules. 
The observed binding energies, molecular alignment, and bond lengths are in good agreement with theory.
Our approach is broadly applicable to Rydberg tweezer platforms, expanding the range of available molecular systems and enabling the integration of Rydberg molecules into existing quantum science platforms.

\end{abstract}
\date{\today}

\maketitle

Understanding and controlling molecules at the quantum level is a fundamental goal in both atomic physics and quantum chemistry, with broad implications for quantum technologies such as quantum simulation, precision measurement, and quantum computation \cite{Cornish2024,DeMille2024}. 
Ultracold molecules provide an excellent platform for investigating molecular phenomena, as their ultralow temperatures enable precise control and delicate probing of quantum states~\cite{Langen2024}. Rydberg molecules, in particular, exhibit several unique features originating from their exaggerated bond lengths which extend to micrometer scales.
This makes them ideal for exploring molecular behavior on macroscopic scales.

Several different types of Rydberg molecules have been experimentally explored, including ultralong-range Rydberg molecules (ULRMs) \cite{Fey2019,eiles2019trilobites,dunning2024ultralong}, where a Rydberg electron binds to a ground-state atom, and Rydberg macrodimers \cite{Shaffer2018,Sasmannshausen2016,Hollerith2019}, which are bound states of two Rydberg atoms. 
While Rydberg molecules have been successfully created in optical lattices \cite{Hollerith2019,Hollerith2021microscopic,Manthey2015}, the overwhelming majority of experimental work, including recent studies of heteronuclear species \cite{Whalen2020,Peper2021}, has been conducted in bulk gases using ion detection methods.  
These methods limit the precise control and detection of individual molecules \cite{peper2023role}.
Such capabilities are highly desirable—and often essential—for realizing theoretical proposals that leverage interactions between Rydberg and ground-state atoms to produce heavy Rydberg states \cite{hummel2020ultracold,peper2020formation}, for quantum information processing \cite{khazali2023scalable}, and to study electron dynamics, including localization and topological behavior, in highly structured microscopic environments \cite{eiles2016ultracold,eiles2023anderson,eiles2024topological}.

Optical tweezer arrays have emerged as a versatile tool for trapping, controlling, and detecting individual atoms with exceptional precision~\cite{Kaufman2021}. 
The interparticle distances in tweezer arrays closely match the bond lengths of Rydberg molecules, offering the possibility of studying these states with single-particle sensitivity.
The recent development of two-species tweezer arrays \cite{Liu2018,Brooks21,Singh2022} introduces independent control over each species, enabling applications such as the assembly of ground-state diatomic molecules \cite{Cairncross2021,Ruttley2024} and the realization of promising new platforms for quantum computing \cite{Beterov2015,Anand2024}. These advancements lay the groundwork for realizing ULRMs in tweezers using a two-species approach.

In this Letter, we demonstrate the creation and study of individual Rb$^{*}$Cs ULRMs using an optical tweezer platform.  By employing single atom detection of Rb and Cs, we observe ULRM formation by correlated loss of both species and leverage the single particle resolution to study the formation dynamics. We characterize the influence of the tweezers on molecule formation, optimizing coupling to the molecular state by manipulating the relative wavefunction of the atom pair. Additionally, we demonstrate the association of molecules with each atom trapped in separate tweezers, allowing precise control over the interatomic distance.

\begin{figure}[] 
\includegraphics[width=\columnwidth]{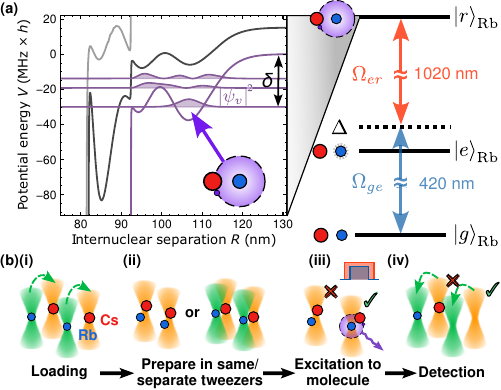}
\caption{(a) Relevant energy levels for excitation of ULRMs. Right: Two-photon Rydberg excitation scheme for Rb atoms. The ground state \ground is coupled to the Rydberg state \ryd via an off-resonant intermediate state \intermediate, with detuning $\Delta$ from $\ket{g}_{\rm Rb}\rightarrow \ket{e}_{\rm Rb}$. Left: Experimentally relevant PECs of the Rb$^{*}$Cs system 
as a function of internuclear separation $R$. The curves are colored according to their asymptotic quantum numbers: $\ket{36\mathrm{S},m_j = -1/2,m_f=3}$ (purple) and $\ket{36\mathrm{S},m_j =1/2,m_f=2}$ (black). The gray PEC does not support bound states in the relevant energy region. Molecular states are supported by these PECs, with vibrational wavefunctions $\psi_{v}$ (purple) shown for $\ket{36\mathrm{S},m_j = -1/2,m_f=3}$. The two-photon detuning $\delta$ is defined with respect to the atomic transition. (b) Experimental scheme for production and detection of Rb$^{*}$Cs ULRMs in tweezers. (i) Rb and Cs atoms are loaded into separate tweezers and Rb atoms are moved towards the Cs tweezers. (ii) Rb and Cs atoms are prepared either in the same tweezer or in separate tweezers. (iii) The atom pair is illuminated with Rydberg light to photoassociate an ULRM. (iv) Rb and Cs are separated into their respective tweezers and detected. Molecule formation results in loss of the atom pair.}
\label{fig:overview}
\end{figure}  
\begin{figure*}[]
\centering
\includegraphics[width=\textwidth]{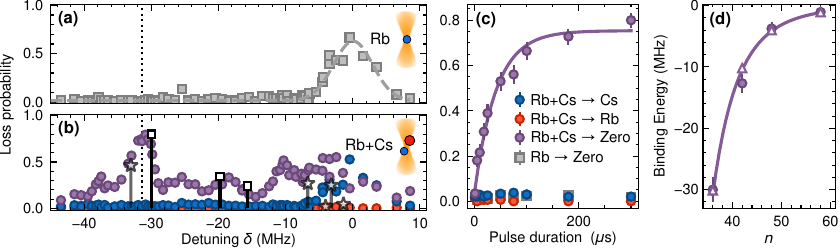}
\caption{
(a,b) Spectroscopy of (a) individual Rb atoms and (b) Rb+Cs pairs in a magnetic field of $4.78$~G and tweezer intensity of $I=93$~kW/cm$^{2}$ as a function of the detuning $\delta$ from the atomic state \Sstate.  
We show the probability of losing an individual Rb atom (gray squares) with a Gaussian fit (gray dashed line) to extract the atomic line position. For atom pairs, the loss probabilities of both atoms (purple), Cs atom (red), or Rb atom (blue) are shown with theoretical predictions for molecular line positions and strengths (black squares). Star-shaped markers with lighter shading indicate bound states that are highly sensitive to the $P$-wave phase shift. (c) Loss probabilities as a function of the duration of the Rydberg pulse with detuning $\delta = -31.4$~MHz (vertical dotted line in (a) and (b)). 
(d) Binding energy of $v=0$ states for $n\mathrm{S}$ Rb$^{*}$Cs ULRMs as a function of principle quantum number $n$. The filled circles (empty triangles) show the measured (predicted) binding energies and the line shows a power-law fit to the data.}
\label{fig:spec}
\end{figure*}
We investigate ULRMs formed from a ground-state $^{133}$Cs atom and a Rydberg $^{87}$Rb atom.
The excited valence electron of the Rb atom interacts strongly with the Cs atom when the internuclear separation $R$ is smaller than the Rydberg orbit. This interaction is characterized by the low-energy $S$- and $P$-wave electron-Cs scattering phase shifts \cite{greene2000creation,fermi1934sopra}. Exemplary Born-Oppenheimer potential energy curves (PECs) are shown in Fig.~\ref{fig:overview}(a).
At large $R$, where the interatomic interactions are negligible compared to the Zeeman splitting, these curves are labeled by the Rydberg electron's magnetic quantum number $m_{j}$ and the quantum number $m_f$ of the ground-state Cs atom; the lower purple (upper black) curve has $m_j = -1/2$ and $m_f=3$ ($m_j =1/2$ and $m_f=2$).
Only the lower curve is experimentally accessible for our choice of initial atomic states; the upper curves are accessible only via state mixing through the electron-Cs interaction.
Molecular states, with binding energy $E_v$ and  vibrational wave functions $\psi_v$, are supported by potential wells in the oscillatory structure of these curves. These oscillations are directly linked to the fluctuating electronic density spatially probed by the Cs atom \cite{greene2000creation}, while the nearly vertical steps at $\sim82$~nm and $92$~nm stem from the low-energy $^3P_J$-wave shape resonances of Cs$^-$ \cite{khuskivadze2002adiabatic, hamilton2002shape,scheer1998experimental}.
The calculation includes both fine and hyperfine structure and the Zeeman interaction with the external magnetic field \cite{eiles2017hamiltonian,hummel2018spin,hummel2019alignment}. The calculation is benchmarked at zero magnetic field using an alternative method based on the Coulomb Green's function \cite{greene2023green}.
Additionally, we use scattering phase shifts incorporating minor adjustments determined by previous measurements of Cs$^-$ photodetachment cross-sections \cite{scheer1998experimental} and Cs$^*$Cs photoassociation spectroscopy \cite{sassmannshausen2015experimental,supplement}.

To prepare ULRMs in tweezers we follow the protocol in Fig.~\ref{fig:overview}(b). Each experiment begins with single Rb and Cs atoms in the motional ground state of species-specific tweezers \cite{supplement}. To prepare the Rb+Cs atom pair at the required separation, we merge a $817$~nm tweezer containing a Rb atom into a 1066~nm tweezer containing a Cs atom. Depending on the experiment, we either maintain the tweezers at a fixed separation or prepare both atoms in the 1066~nm tweezer by adiabatically reducing the 817~nm tweezer intensity. Using the two-photon excitation scheme shown in  Fig.~\ref{fig:overview}(a), we excite the Rb+Cs atom pair to a Rb$^{*}$Cs ULRM. After excitation, we separate any remaining atoms into their respective tweezers. We take atomic fluorescence images at the start and end of each sequence to determine each tweezer's occupancy.

Figure~\ref{fig:spec} shows spectroscopy of Rb*Cs ULRM in tweezers. We first investigate molecular states below the atomic state \Sstate (PEC shown in Fig.~\ref{fig:overview}), with the Rb+Cs atom pair prepared in the same 1066~nm tweezer. We scan the Rydberg laser frequency below the atomic state and measure the probability of losing a Rb or Cs atom. Non-destructive imaging of Rb and Cs atoms in separate tweezers before excitation allows postselection for cases with only a Rb atom  [Fig.~\ref{fig:spec}(a)] or a Rb+Cs pair [Fig.~\ref{fig:spec}(b)] initially present. 

For the Rb-only case, we observe a single loss peak corresponding to the atomic line. This loss occurs because Rydberg atoms are anti-trapped at our tweezer wavelength. The contrast of this peak is limited by the likelihood that a Rydberg atom decays before escaping the tweezer; most decay events populate another hyperfine state of the ground manifold, which does not couple to our Rydberg lasers.  The $150 \, \mu$s excitation pulse saturates the atomic transition, leading to significant broadening.

In Fig.~\ref{fig:spec}(b) the blue markers show events where we detect only the loss of the Rb atom and keep the Cs atom, showing a similar peak around the atomic line. Additionally, the purple markers reveal extra peaks corresponding to the loss of both atoms from the tweezers, indicating Rb*Cs ULRM formation. 
We assign states to molecular peaks by comparing their positions to theoretical calculations \cite{supplement}. The stick spectrum locations indicate theoretical line positions, with heights showing expected line strengths. 
These strengths incorporate electronic transition strengths and vibrational Franck-Condon factors (FCFs).
We assign the strongest peak around $\delta=-31$~MHz as the $v=0$ state. The black lines correspond to the three vibrational levels in Fig.~\ref{fig:overview}(a), whose energies are insensitive to the $^3P_0$ shape resonance position. In contrast, the lighter shading indicates states whose energies vary significantly within the uncertainty range of the measured resonance position in Ref.~\cite{scheer1998experimental}.

Exact counting of the remaining particles after Rydberg excitation reveals the loss dynamics. In Fig.~\ref{fig:spec}(c) we tune the laser to the $v=0$ peak and measure particle loss over time. 
We observe significant pair loss but no apparent single-particle loss, suggesting that molecule loss involves a process ejecting both Rb and Cs from the tweezer, likely due to radiative decay. This decay converts molecular binding energy into kinetic energy which exceeds the tweezer depth, causing pair loss. Additionally, we observe that the lineshape of the $v=0$ feature in Fig.~\ref{fig:spec}(b) is asymmetric and significantly broader than expected \cite{supplement}. The cause remains unclear but persists for all molecular resonances observed at $4.78$~G in the same tweezer.

In Fig.~\ref{fig:spec}(d) we show the measured binding energies of $v=0$ states for all $n\mathrm{S}$ ULRMs explored in this work. 
The empty triangles show theoretical predictions \cite{supplement} which are in excellent agreement with the measured binding energies. This agreement spans more than one order of magnitude in energy and is achieved without fitting any parameters to our experimental data. 

\begin{figure}[] 
\includegraphics[width=\columnwidth]{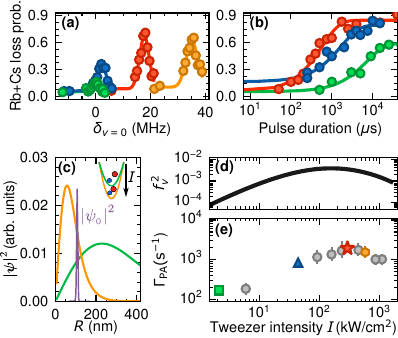}
\caption{(a) Probability of losing a Rb+Cs atom pair as a function of the two-photon detuning from the molecular state $v=0$ at zero tweezer intensity $\delta_{v=0}$. The different colors represent spectra obtained at different tweezer intensities 2.2~kW/cm$^{2}$ (green), 45~kW/cm$^{2}$ (blue), 297~kW/cm$^{2}$ (red) and 586~kW/cm$^{2}$ (orange)  respectively. (b) Atom pair loss as a function of Rydberg pulse duration on resonance with the molecular state. The colors correspond to the same tweezer intensities as in (a) and the solid lines are fits to the experimental data \cite{supplement}. (c) Probability density for the $v=0$ molecular state $\left|\psi_{0}\right|^{2}$ (purple) and the ground state atom pair (green and orange). The atom pair wavefunctions show the projection of the cylindrical wavefunction onto the $\tilde l = 0,m_{\tilde l}=0$ spherical harmonic \cite{supplement}, with colors corresponding to the same tweezer intensities as above. (d) The black line shows $f_{v}^{2}$, the Franck-Condon factor squared, as a function of the tweezer intensity $I$. 
(e) Measured atom pair loss rates $\Gamma_{\rm PA}$ as a function of the tweezer intensity. Colored markers correspond to the same tweezer intensities as in (a); gray markers show additional intensities omitted in (a) for clarity.}

\label{fig:FCF}
\end{figure}

Next, we investigate the influence of the tweezers on the formation of ULRMs. Figure~\ref{fig:FCF} shows how the tweezer intensity affects molecules in the ground vibrational state $v=0$ of the PEC which approaches \Sstate asymptotically.
In Fig.~\ref{fig:FCF}(a) we observe that higher tweezer intensity increases the transition frequency. 
This transition shift results from the differential ac Stark shift $\propto (\alpha_{\rm Rb+Cs}-\alpha_{v})I$, where $\alpha_{\rm Rb+Cs}$ and $\alpha_{v}$ are the polarizability of the atom-pair in \groundpair and the ULRM in state $v$ respectively. This measured shift is identical to that measured for the atomic transition \ground $\rightarrow$ \ryd \cite{supplement}, indicating that tweezer intensity does not affect the ULRM's binding energy. Therefore, the molecular polarizability is the sum of the polarizabilities of the constituent particles, $\alpha_{v}=\alpha_{\rm Cs}+\alpha_{r}$.
At our tweezer wavelength, this implies a polarizability of $\alpha_{v}= 613\times4\pi\varepsilon_0 a_0^3$,  indicating the ULRMs are trapped.

The tweezer intensity also alters the probability of exciting the atoms to the molecular state, as seen by the changing peak heights in Fig.~\ref{fig:FCF}(a) for a fixed Rydberg pulse duration. Figure~\ref{fig:FCF}(b) shows pair loss as a function of pulse duration for three different intensities, with the laser tuned to the shifted molecular transition for each intensity. We fit the data to an exponential model, extracting the loss rate $\Gamma_{\rm PA}$ \cite{supplement}.
The changing loss rate as a function of tweezer intensity can be understood by considering the FCF $f_v$ that dictates the coupling to the molecular state. The FCF is determined by the wavefunction overlap between the molecule and atom pair. 
In Fig.~\ref{fig:FCF}(c), we plot the radial probability density of the rovibrational ground state $v=0$, which is strongly peaked around 100~nm. The green and orange curves depict the isotropic component of the wavefunction for an atom-pair trapped in an anisotropic tweezer potential \cite{supplement}. Increasing $I$ from 2.2~kW/cm$^{2}$ (green) to 586~kW/cm$^{2}$ (orange) shifts the peak density to smaller $R$ while narrowing its spread. 
In Figure~\ref{fig:FCF}(d) we plot $f_v^{2}$ for our experimental parameters. The predicted FCF peaks at $6\%$ around an intensity of 200~kW/cm$^{2}$. For larger intensities, we can reach the regime where the atom pair's most likely separation shifts to distances smaller than the molecular bond length, as shown by the orange curve in Fig.~\ref{fig:FCF}(c). In Fig.~\ref{fig:FCF}(e) we plot the loss rate extracted from the measurements in Fig.~\ref{fig:FCF}(b). As the coupling to the molecular state is $\Omega_{v} \propto f_{v}$, we expect the measured atom pair loss rate is $\Gamma_{\rm PA} \propto f_{v}^{2}$ \cite{supplement}. The experimental data matches the predicted behaviour well and we observe an order of magnitude increase in the loss rate for intensities between $200-600$~kW/cm$^{2}$.

Controlling the FCF between the atom pair and ULRMs using the trap intensity provides a valuable experimental tool, previously limited to optical lattice experiments \cite{Thomas2018b,Hollerith2019}. However, by trapping the atoms in the same tweezer we cannot decouple the most likely separation of the atom pair from the wavefunction spread. A key feature of the tweezer platform is the ability to reconfigure the array geometry, which we now exploit to position the atom pair at the separation required to maximize the FCF.

\begin{figure}[] 
\includegraphics[width=\columnwidth]{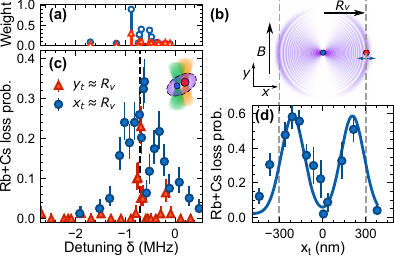}
\caption{Assembly of ULRMs in separate tweezers. (a) Blue circles (red triangles) show theoretical predictions for line strengths and positions for tweezers separated by the bond length $R_{v}=300$~nm  along the x-axis (y-axis). (b) The electron probability density of the state \Dstate. A 4.78~G magnetic field $B$ is applied along the y-axis. The dashed lines show the calculated bond length $R_{v}=300$~nm of the state $v=0$. (c) Pair loss probability as a function of detuning $\delta$ from the atomic state \Dstate. The Rb and Cs atoms are trapped in distinct tweezers (wavelengths 817~nm  and 1066~nm respectively), separated by 300~nm along x (blue) or y (red). (d) Pair loss probability as a function of the separation x$_{t}$ of the 817~nm and 1066~nm tweezers along the x-axis. 
The two-photon detuning is resonant with the loss feature indicated by the black dashed line in (c) and 
the solid blue line shows a double Gaussian fit to the experimental data. }
\label{fig:sep}
\end{figure}

We demonstrate the preparation of an ULRM across two separate tweezers in Fig.~\ref{fig:sep}. Here, we retain Rb in the 817~nm tweezer. The relative separation of the two tweezers is tuneable in all three dimensions, with a positional resolution of approximately 50 nm in the x-y plane [Fig.~\ref{fig:sep}(b)] and 100 nm in the x-z plane, limited by shot-to-shot fluctuations in the relative position of the tweezers~\cite{Guttridge2023}. To increase the bond length $R_{v}$ relative to this positional fluctuation, we explore ULRMs at higher $n$. 
Figure~\ref{fig:sep}(b) shows the electron probability density for the atomic state \Dstate. We utilise the anisotropic wavefunction of this D state to demonstrate the control over the constituent particles afforded by our approach. The PEC's depth reduces to zero when the atoms are oriented along the quantization axis \cite{supplement}, meaning there are no molecular bound states. Figure~\ref{fig:sep}(a) shows theoretical predictions for line strengths and positions for tweezers separated by the bond length $R_{v}=300$~nm  along the x-axis (blue circles) and y-axis (red triangles).  We predict the ratio of the FCFs between the parallel and perpendicular alignment is non-zero and approximately $0.3$ \cite{supplement}, caused by the residual atom-pair's wavefunction spread in the x-z plane.

Figure~\ref{fig:sep}(c) shows spectroscopy of ULRMs below the atomic state \Dstate. The blue (red) markers show data obtained with the tweezers separated by approximately the bond length and oriented perpendicular (parallel) to the applied magnetic field. The measured spectrum exhibits a clear contrast between the two orientations and is in good agreement with our theoretical predictions.

In Fig.~\ref{fig:sep}(d) we investigate how the coupling to the molecular state depends on the tweezer separation. We fix the detuning on resonance with the molecular line (dashed line in Fig.~\ref{fig:sep}(c)) and orient the atom pair along the x-axis, perpendicular to both the quantization axis and the tweezer propagation direction. We observe a striking dependence of atom loss on tweezer position, with a peak in the loss occurring when the Cs atom is positioned on either side of the Rb atom. Fitting the data with a double Gaussian, we extract an average displacement from the center of $210(10)$~nm. To compare with the expected bond length, we must account for the difference between the \emph{tweezer} separation and the \emph{atomic} separation, arising from the distinct atomic polarizabilities. For instance, the Cs atom is repelled by the 817~nm tweezer, increasing the atomic separation relative to the tweezer separation~\cite{Ruttley2023}. Numerical simulations of the tweezer potential~\cite{supplement} indicate that a tweezer separation of $\mathrm{x}_{t}=210$~nm corresponds to an atomic separation of $R=300$~nm, consistent with the expected bond length.

Our demonstrated manipulation of the initial atom-pair state is promising for the realization of other forms of ULRMs such as giant polyatomic Rydberg molecules \cite{Rittenhouse2010}, which could be realized by combing diatomic molecules with Rydberg atoms \cite{GonzalezFerez2020,MelladoAlcedo2024}. Trapping of the constituent particles in separate tweezers bypasses problems with preparing atoms and molecules in the same trap~\cite{Gregory2021} and can enhance the coupling to molecular states. This approach enables the selective preparation of different molecular symmetry groups and exotic molecular states based on the initial atom geometry \cite{eiles2016ultracold}, and opens up the possibility of manipulating ULRMs using tweezers. Importantly, our method is readily extendable to homonuclear systems \cite{supplement}, facilitating studies of both well-established and novel molecular species.

In conclusion, we have presented a new approach to studying ULRMs using optical tweezers, reporting the formation and characterization of individual Rb$^{*}$Cs ULRMs. By leveraging the inherent single-particle manipulation of our approach, we have demonstrated control over the atom pair's relative wavefunction, optimizing the coupling to molecular states and achieving molecular association with atoms trapped in separate tweezers.

{Our approach opens new pathways for exploring polyatomic ULRMs in programmable geometries and enables trapping of ULRMs for coherent control. Moreover, extending this method to other Rydberg tweezer platforms will expand the range of molecular species that can be studied and create new opportunities for integrating ULRMs into quantum simulation and computation platforms.}

The data that support the findings of this article are openly available \cite{dataset}.

\begin{acknowledgments}
We thank M. P. A. Jones for helpful discussions. AD and ME thank F. Hummel for assistance with the magnetic field interaction. We acknowledge support from UK Research and Innovation (UKRI) Frontier Research Grant EP/X023354/1, the Royal Society and Durham University.

\end{acknowledgments}

A.G. and M.T.E. conceptualized the work; A.G. and T.R.H. performed the experiments; D.K.R. contributed to the experimental apparatus; A.A.T.D. and M.T.E. performed the theoretical work; S.L.C. supervised the experimental work. A.G. wrote the manuscript with contributions from all authors.

%

\ifarXiv
    

\section*{Supplementary material}

# Supplemental Material: Individual assembly of two-species Rydberg molecules using optical tweezers

Alexander Guttridge,[1, 2, *] Tom R. Hepworth,[1, 2] Daniel K. Ruttley,[1, 2] Aileen A. T. Durst,[3] Matthew T. Eiles,[3] and Simon L. Cornish[1, 2]

[1]*Department of Physics, Durham University, South Road, Durham, DH1 3LE, United Kingdom*
[2]*Joint Quantum Centre Durham-Newcastle, Durham University, South Road, Durham, DH1 3LE, United Kingdom*
[3]*Max Planck Institute for the Physics of Complex Systems, Nöthnitzer Str. 38, 01187 Dresden, Germany*

## THEORETICAL DESCRIPTION OF RB*CS RYDBERG MOLECULES

In this section, we describe the properties of a long-range Rb*Cs Rydberg molecule in a magnetic field, providing details of our general theoretical approach and specifying the simpler model used to describe the $n$D-state Rydberg molecules. We use atomic units throughout all of this section.

### Molecular Hamiltonian

The Rydberg molecules studied in this work are described by the molecular Hamiltonian $\hat{H}$, where

$$\hat{H} = \hat{H}_{\text{nuc}} + \hat{H}_{\text{Ryd}} + \hat{V}_{e-\text{Cs}} + \hat{H}_{\text{HF}} + \hat{H}_B. \quad (S1)$$

In the relative coordinate $\vec{R}$ of the two nuclei, the Hamiltonian involving only the nuclear degrees of freedom is

$$\hat{H}_{\text{nuc}} = -\frac{\nabla_R^2}{2\mu_{\text{red}}} - \frac{\alpha_{\text{Cs}}}{2R^4}, \quad (S2)$$

where $\mu_{\text{red}} = \frac{m_{\text{Rb}} m_{\text{Cs}}}{m_{\text{Rb}} + m_{\text{Cs}}}$ is the reduced mass and $\alpha_{\text{Cs}}$ is the polarizability of the Cs atom. The second term of $\hat{H}_{\text{nuc}}$ is the long-range polarization potential between the positively charged core of the Rydberg Rb atom and the ground-state Cs atom.

The Hamiltonian of the Rydberg atom, assuming an infinitely heavy Rb core at $r = 0$, is

$$\hat{H}_{\text{Ryd}} = -\frac{\nabla_r^2}{2} - \frac{1}{r} + \hat{V}_{\text{sr}}^{l,j}(r), \quad (S3)$$

where $\hat{V}_{\text{sr}}^{l,j}(r)$ describes the non-Coulombic interaction between the Rydberg electron and the Rb$^+$ core at small $r$. The effect of this interaction is encapsulated in the fine structure-dependent quantum defects $\mu_{lj}$. The eigenstates and eigenenergies of $\hat{H}_{\text{Ryd}}$, $|n(s_1 l)jm_j\rangle$ and $E_{n(sl)jm_j} = -\frac{1}{2(n-\mu_{lj})^2}$, respectively, are well-known [S1, S2].

The interaction between the Rydberg electron and the Cs atom, $\hat{V}_{e-\text{Cs}}$, is accurately described by the pseudopotential first introduced by Fermi (Ref. [S3]) and later extended by Omont (Ref. [S4]) for $S$- and $P$-wave scattering partial waves, respectively. Relativistic effects play an important role in Cs, so it is essential to include both singlet and triplet scattering channels as well as the spin-orbit splitting. The calculated electron-Cs phase shifts $\delta_{(LS)J}(k)$ are therefore dependent on the $^{2S+1}L_J$ term symbol of the Cs anion. The energy-dependent phase shifts are evaluated using the semiclassical momentum of the electron having energy $E_e$, $k = \sqrt{2(E_e + 1/R)}$. The resulting pseudopotential has the structure [S5]

$$\hat{V}_{e-\text{Cs}}(\vec{r}, \vec{R}) = -\sum_{(LS)J} |(LS)JM_J\rangle \frac{(2L+1)^2}{2} \quad (S4)$$

$$\times \frac{\tan\left(\delta_{(LS)J}(k)\right)}{(kX)^{2L+1}} \frac{\delta(X)}{X} \langle (LS)JM_J|.$$

Here, $X = |\vec{r} - \vec{R}|$ is the distance between the Rydberg electron and the Cs atom. We include contributions from six total scattering partial waves: $^1S_0$, $^3S_1$, $^1P_1$, and $^3P_{0,1,2}$.

The hyperfine Hamiltonian is

$$\hat{H}_{\text{HF}} = A \vec{i}_{\text{Cs}} \cdot \vec{s}_2, \quad (S5)$$

where $A = 2.298$ GHz for $^{133}$Cs, which has nuclear spin $i_{\text{Cs}} = 7/2$ and electron spin $s_2 = 1/2$. This term is diagonal in the coupled spin basis of the perturber atom $|(i_{\text{Cs}} s_2) f m_f\rangle$, where $f$ is the Cs atom's total spin and $m_f$ is the projection of this spin along the quantization axis. The hyperfine structure of the Rydberg Rb atom is negligible.

Finally, as an external magnetic field is always present in our experiment, we include the magnetic Hamiltonian $\hat{H}_B$. The interaction between the Rydberg molecule and this field is determined by the linear Zeeman term, which couples the magnetic field to the total angular momentum of the Rydberg atom, its electronic spin, and the electronic spin of the ground-state atom:

$$\hat{H}_B = \frac{\vec{B}}{2} \cdot (\vec{j} + \vec{s}_1 + 2\vec{s}_2). \quad (S6)$$

In our modeling of the experimental results for a magnetic field of 4.78 G, we neglect the diamagnetic term, whose strength is proportional to $B^2 n^4$. For the highest principal quantum number studied here, $n = 56$, the effect of this term does not exceed 50 kHz. However, for the larger field of $\approx 182$ G applied in some of the measurements, the strength of the diamagnetic term ranges

from 10 to 80 MHz as $n$ increases from 33 to 56. This shift will further separate molecular potential curves but not modify their structure, so unless a degeneracy is coincidentally created it should not have a noticeable effect.

### Adiabatic potential energy surfaces

The latter four terms of $\hat{H}$ (Eq. S1) make up the electronic Hamiltonian $\hat{H}_e$. The eigenvalues $U_\mu(R,\theta)$ and eigenstates $\phi_\mu(\vec{r};R,\theta) = \sum_\alpha C_{\mu\alpha}(R,\theta)\langle\vec{r}|\alpha\rangle$ of this Hamiltonian, obtained at fixed $\vec{R}$, are the Born-Oppenheimer potential energy surfaces and electronic eigenstates, respectively. We align the quantization axis with the magnetic field, so that these are functions of the internuclear distance $R$ separating the two atoms and the angle $\theta$ between the molecular axis and the magnetic field. As a basis, the product states of the Rydberg atom and the total spin state of the Cs ground-state atom are used:

$$|\alpha\rangle = |n(ls_1)jm_j\rangle \otimes |(i_{Cs}s_2)fm_f\rangle. \quad (S7)$$

Further details of the electronic Hamiltonian, its matrix elements in this basis, and its diagonalization are given in Ref. [S5] for the field-free case and Ref. [S6, S7] for the case of an applied magnetic field. All spin states and allowed angular momenta are included in the basis, which is truncated to finite size by including a subset of Rydberg states $n$ surrounding the level of interest.

Despite this truncation, the size of the basis becomes unwieldy due to the high degeneracy of participating Rydberg levels at an arbitrary field geometry. For example, for a fixed $n$ and $l$, the number of states is $\{[2(l+1/2)+1]+[2(l-1/2)+1]\}\cdot\{[2(i_{Cs}+1/2)+1]\cdot[2(i_{Cs}-1/2)+1]\} = 4(2l+1)(2i_{Cs}+2)$. For $l=2$, this results in subspace consisting of 160 states; for all $l$ levels included in a fixed $n$ (say, $n=56$), this is nearly 100,000 states. We therefore perform a full diagonalization only for the aligned case with $\theta=0$. Here, $M_{tot}=m_j+m_f$ is a good quantum number. Further, since Rydberg states with $m_j > 3/2$ are not perturbed by the Cs atom with this alignment, $M_{tot}$ for the Rb*Cs molecular system cannot exceed the value 11/2. By restricting the basis to states with a fixed $M_{tot}$, we limit the basis to a few thousand states. For the Rydberg $n$S states studied here, this assumption that the molecular potential is independent of $\theta$ is an excellent approximation, particularly so at large $R$ where the most localized molecular states form [S7]. Therefore, for the rest of this section, it will be assumed that we discuss the Rydberg $n$S states and compute only spherically symmetric molecular potentials. The Rydberg $n$D states must be treated with a different approach, as the angular dependence of the potential surface plays an important role. We discuss the treatment of these states later, where we describe how to obtain the relevant

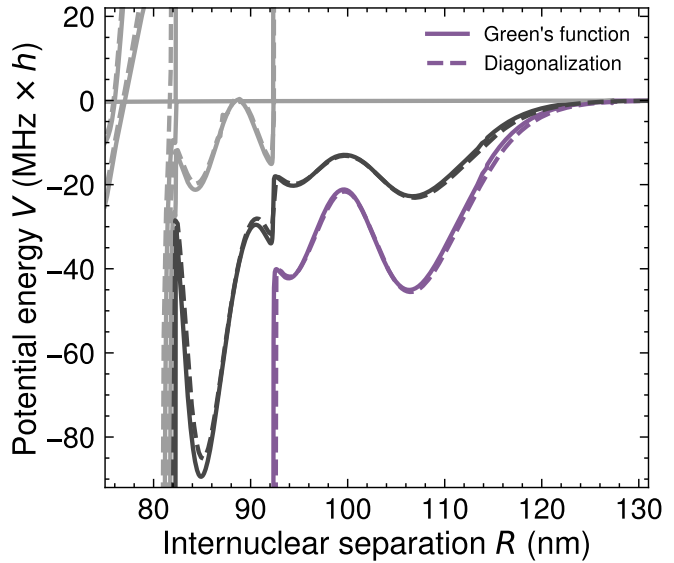

FIG. S1. Comparison of the potential energy curves for the 36S Rb*Cs Rydberg molecule shown in Fig. 1 of the main text. These were computed by directly diagonalizing the electronic Hamiltonian in a truncated basis set (dashed) and via the Green's function method (solid). In the relevant range of $R$ – the outermost potential well – the methods agree to better than 2%.

potential energy surfaces in a more tractable perturbative limit, and how we solve for aligned molecular states and relevant Franck-Condon factors in that regime.

### Numerical accuracy

As discussed in Refs. [S1, S8–S10], it is not possible to rigorously converge the potential energy surfaces with respect to the size of the basis. This is caused by the singular nature of the Fermi pseudopotential, the existence of scattering shape resonances in the $P$-wave channel which cause the tangent function in Eq. S4 to diverge, and the ambiguity in the choice of "reference" electron energy $E_e$ in the definition of $k$.

A Green's function-based approach, recently extended to the fully spin-dependent system [S10], circumvents these three problems. Formally, the basis set is no longer truncated due to the inclusion of all Rydberg states in the Coulomb Green's function, and as this method relies on the scattering phases rather than the scattering lengths or scattering volumes, the $P$-wave shape resonances cause no difficulty. Finally, the electronic energy $E_e$ is determined self-consistently in this method, removing the third theoretical hurdle.

However, when an external field is present the Coulomb Green's function, the backbone of this method, is no longer the appropriate choice. We therefore resort to using this method to benchmark the diagonalization

method described above in the zero-field limit. An example benchmark is shown in Fig. S1 for the 36S Rb*Cs potential curves. The potential curves from the diagonalization were computed using a basis including all Rydberg levels with principal quantum numbers $31 < n \leq 33$. Additionally, the energy of the unperturbed $n$S Rydberg state was used as the electron energy in the definition of $k$. The positions and depths of the outermost well are in excellent agreement, as is the position of the collection of sharp avoided crossings at $R \sim 92$ nm caused by level repulsion from the "butterfly" potential energy curve. As $R$ decreases, the disagreement between methods grows due to the increasingly large $P$-wave interaction. Similar agreement between these methods is obtained for the other Rydberg states considered here, and is somewhat fortuitous - in the similar calculations of $n$P Cs*Cs molecules discussed below, the potential curve depths differed by about 10%.

### Scattering phase shifts

Even with an accurate and benchmarked method to calculate the potential energy curves, the results are only as reliable as the atomic parameters used for the various interactions. The Rb quantum defects and Cs hyperfine splitting are determined to high precision by experiment [S2], but no high-resolution experimental data exist for the low-energy scattering phase shifts. These have been calculated using effective range theory and by direct solution of the Dirac equation [S11, S12], but key parameters such as the zero-energy $S$-wave scattering length or $^3P_J$ shape resonance positions differ by 10% or more in different calculations.

Ref. [S13] proposed that the triplet $S$-wave scattering length of Cs is about 13% smaller than its literature value of $-22.8 \, a_0$ [S12, S14]. This is based on a comparison of the measured vibrational spectra of K*Cs with those calculated using these modified phase shifts and the diagonalization method. However, it could not be ascertained how much of this change in the scattering length is physical, and how much arose from non-converged potential energy curves. We therefore applied the Green's function approach to compute potential energy curves for homonuclear Cs$_2$ ultralong-range Rydberg molecules. By comparing the vibrational spectra computed for these potential curves with the measurements of Ref. [S15], we could obtain an independent verification of this correction to the theoretical scattering length. We found that a 10% reduction was necessary to guarantee excellent agreement in the positions of the strongest resonances across the range of principal quantum numbers observed there. Table I shows measured binding energies from Ref. [S15] and our calculated values for the extremal principal quantum numbers reported. Finally, to take into account the measured position of the $^3P_1$ scattering resonance [S16], we modified all three $^3P_J$ phase shifts such that their resonance positions shifted upwards in energy by about 1.3 meV. This did not change the positions of the binding energies reported in Table I, as these states are bound in the outermost potential well which is predominantly $S$-wave.

| State | Exp. (MHz) | Th. (MHz) |
|---|---|---|
| $26P_{3/2}, f = 3$ | -406 | -407 |
| $26P_{3/2}, f = 3$ | -209 | -206 |
| $26P_{3/2}, f = 4$ | -409 | -408 |
| $26P_{3/2}, f = 4$ | -144 | -146 |
| $33P_{3/2}, f = 3$ | -90 | -90 |
| $33P_{3/2}, f = 3$ | -51 | -51 |
| $33P_{3/2}, f = 4$ | -89 | -88 |
| $33P_{3/2}, f = 4$ | -35 | -33 |

TABLE I. Cs*Cs Rydberg molecule binding energies reported in Ref. [S17] (Exp.) and computed with the Green's function method (Th.). The reported states are the $v = 0$ ground state in the "pure triplet" (deeper binding energy) and "mixed" (shallower binding energy) potential curves for Rydberg molecules with two different Rydberg states and two different hyperfine levels.

### Evaluation of binding energies and vibrational wave functions

Once the adiabatic potential energy curves $U_\mu(R)$ are obtained, the radial Schrödinger equation for the nuclear degrees of freedom

$$\left(-\frac{1}{2\mu_{\text{red}}}\frac{d^2}{dR^2} + \frac{\tilde{l}(\tilde{l}+1)}{2\mu_{\text{red}}R^2} + U_\mu(R) - E_{v\tilde{l}}\right)\psi^\mu_{v\tilde{l}}(R) = 0 \quad \text{(S8)}$$

must be solved. Here, $\tilde{l}$ is the rotational angular momentum of the molecule, and $\psi^\mu_{v\tilde{l}}(R)$ is the molecular state with vibrational quantum number $v$, angular momentum $\tilde{l}$, residing in the $\mu$th adiabatic potential energy curve. Due to the enormous bond length, the rotational splitting is on the order of kHz and not resolved experimentally.

Because of the $P$-wave scattering shape resonance, the adiabatic potentials $U_\mu(R)$ exhibit only a handful of shallow potential wells, with depths ranging from 1-100 MHz for $n$ varying from 33 to 56, before plunging steeply down several tens of GHz to the next atomic line, which it intersects at a distance $R_c$. Therefore, although the nuclear wave function vanishes as usual at large $R$, $\psi^\mu_{v\tilde{l}}(R \to \infty) = 0$, we solve Eq. S8 subject to the Siegert boundary condition

$$\partial\psi^\mu_{v\tilde{l}}(R)/\partial R|_{R=R_0} = -i\sqrt{2\mu_{\text{red}}E_{v\tilde{l}}}\,\psi^\mu_{v\tilde{l}}(R = R_0) \quad \text{(S9)}$$

at a point $R_0 \ll R_c$. This boundary condition allows us to characterize the molecular states as scattering reso-

nances, and calculate their positions and widths via diagonalization of a quadratic eigenvalue equation [S18]. More details are given in Ref. [S19]. In most cases, all molecular states except for the $v = 0$ state bound in the outer-most well lack an internal potential barrier and have finite lifetimes. These states emerge as resonances "bound by quantum reflection" [S20]. Recent work has suggested that the lifetimes of these states can be significantly extended by non-adiabatic coupling and their binding energies can be modified, even up to 10% [S19].

### Rydberg $n$S state conclusions

In conclusion, to obtain the Born-Oppenheimer potential energy curves we diagonalized the electronic Hamiltonian. The basis used to describe this Hamiltonian was truncated so that the resulting potential energy curves agreed with those from a more accurate Green's function method in the field-free case. The scattering phase shifts were modified following Ref. [S21], and were verified to produce good agreement with the experiment reported in [S15]. We used these same parameters and basis construction for all four $n$S Rb*Cs molecules observed in the present experiment. With them, we obtained good agreement for all observed $v = 0$ states (see Fig. 2d of the main text).

## RYDBERG D STATE MOLECULES IN A MAGNETIC FIELD

As discussed earlier, Rydberg $n$D states require a different, simplified, treatment as their potential surfaces depend on the angle $\theta$ between the magnetic field direction and the internuclear axis. This makes the full calculation described in the previous section intractable, as the selection rule on $M_{\text{tot}}$ is broken and the huge degeneracy of Rydberg levels makes the basis size enormous. Furthermore, the potentials must be evaluated on a two-dimensional spatial grid. For these reasons, we develop in this section two different perturbative models which can be used to obtain the key qualitative features (structure of the potential curves, anisotropy, and overall strength of the interaction) without solving the entire problem. This calculation is similar to that performed in Refs. [S22, S23].

### Perturbative model: diagonalization in a reduced subspace

The idea here is to obtain the potential energy curves in a perturbative limit where the principal quantum number $n = 56$ and orbital angular momentum $l = 2$ are fixed, but all other degrees of freedom are included in the basis. This approximation should be reasonably accurate as the energetic separation between states with different $l$ is larger than the hyperfine splitting, the magnetic field interaction, the fine structure splitting, or the interaction with the ground state atom. As before, we use the basis states in the representation where the total angular momenta $j$ and $f$ of the two atoms are kept uncoupled,

$$|\alpha\rangle = |n = 56, (l = 2, s_1)jm_j, (i_{\text{Cs}} s_2)fm_f\rangle. \quad \text{(S10)}$$

In this basis, we diagonalize the electronic Hamiltonian

$$\hat{H}_e = \hat{H}_{\text{Ryd}} + \hat{V}_{e-\text{Cs}} + \hat{H}_{\text{HF}} + \hat{H}_B. \quad \text{(S11)}$$

In this representation, the Hamiltonians describing the two isolated atoms are diagonal:

$$\langle \alpha | \hat{H}_{\text{Ryd}} | \alpha' \rangle = \delta_{\alpha\alpha'} E_j \quad \text{(S12)}$$
$$\langle \alpha | \hat{H}_{\text{HF}} | \alpha' \rangle = \delta_{\alpha\alpha'} E_f. \quad \text{(S13)}$$

However, the magnetic field term is not, and additionally requires us to uncouple the electronic spins before evaluating. Its matrix elements are:

$$\begin{aligned}
\langle \alpha | \hat{H}_B | \alpha' \rangle &= \sum_{m,m_1} \sum_{m',m_1'} \sum_{m_i,m_2} \sum_{m_i',m_2'} C_{lm,s_1m_1}^{jm_j} C_{im_i,s_2m_2}^{fm_f} C_{lm',s_1m_1'}^{j'm_j'} C_{im_i',s_2m_2'}^{f'm_f'} \\
&\quad \times \left\langle nlm, s_1m_1, im_i, s_2m_2 \left| \frac{B}{2} \left(2\hat{s}_{1z} + 2\hat{s}_{2z} + \hat{l}_z\right) \right| nlm', s_1m_1', im_i', s_2m_2' \right\rangle \\
&= \frac{B}{2} \delta_{m_j m_j'} \delta_{m_f m_f'} \sum_{m,m_1} \sum_{m_i,m_2} C_{lm,s_1m_1}^{jm_j} C_{im_i,s_2m_2}^{fm_f} C_{lm,s_1m_1}^{j'm_j} C_{im_i,s_2m_2}^{f'm_f} (2m_1 + 2m_2 + m).
\end{aligned} \quad \text{(S14)}$$

Finally, the electron-Cs interaction term is also not diagonal, and its matrix elements require an additional coupling of the spins together to form singlet and triplet states (note that in the following expressions we denote $i_{\text{Cs}}$ simply by

$i$ for brevity):

$$\langle\alpha|\hat{V}_{e-\text{Cs}}|\alpha'\rangle = \sum_{m,m_1}\sum_{m',m'_1}\sum_{m_i,m_2}\sum_{m'_i,m'_2}\sum_{SM_S}\sum_{S'M'_S} C^{jm_j}_{lm,s_1m_1} C^{fm_f}_{im_i,s_2m_2} C^{j'm'_j}_{lm',s_1m'_1} C^{f'm'_f}_{im'_i,s_2m'_2} C^{SM_S}_{s_1m_1,s_2m_2} C^{S'M'_S}_{s_1m'_1,s_2m'_2}$$

$$\times \langle nlm, SM_S, im_i| \left( \sum_{TM_T} |TM_T\rangle 2\pi\delta^3(\vec{r}-\vec{R}) [a_T^s + 3a_T^p \nabla\cdot\nabla] \langle TM_T| \right) |nlm', S'M'_S, im'_i\rangle$$

$$= 2\pi \sum_{m,m_1}\sum_{m',m'_1}\sum_{m_i,m_2}\sum_{m'_2}\sum_{SM_S} C^{jm_j}_{lm,s_1m_1} C^{fm_f}_{im_i,s_2m_2} C^{j'm'_j}_{lm',s_1m'_1} C^{f'm'_f}_{im'_i,s_2m'_2} C^{SM_S}_{s_1m_1,s_2m_2} C^{S'M'_S}_{s_1m'_1,s_2m'_2}$$

$$\times \left[ a_T^s \phi_{nl}^*(R)\phi_{nl}(R) + 3a_T^p \frac{d\phi_{nl}^*(R)}{dR}\frac{d\phi_{nl}(R)}{dR} \right] Y_{lm}^*(\theta,0) Y_{lm'}(\theta,0). \tag{S15}$$

Here, $\phi_{nl}(R)$ is the radial hydrogenic wave function (for simplicity, we neglect the inconsequential $j$-dependence of the quantum defect here) and the spherical harmonics are evaluated at $\varphi=0$ since cylindrical symmetry guarantees the results are isotropic in $\varphi$. The radial-dependent term in square brackets in the final line will be referred to as $y_{nT}(R)$ in further expressions. Here, we have ignored the spin-orbit coupling of the scattering interaction, which splits the $P$-wave interaction into three different resonances. Finally, we also ignore the angular derivatives in the $P$-wave term of the pseudopotential, as these do not couple to terms with predominantly $S$-wave scattering and are therefore not relevant to the states we want to investigate.

We diagonalize the $160 \times 160$ matrix representation of $\hat{H}_e$ (Eq. S11) in this basis to obtain the potential energy surfaces for the desired 56D states.

### Perturbative model: single matrix elements

To qualitatively understand the $\theta$-dependence of these potential surfaces, we also utilize a treatment which considers only the diagonal elements of this matrix. This yields analytic forms for the potential surfaces and involves evaluation of only four diagonal matrix elements $V^{f,m_f}_{j,m_j}$, where for example

$$V^{3,3}_{5/2,1/2}(R,\theta) = \left\langle n\left(2\frac{1}{2}\right)\frac{5}{2}\frac{1}{2},33 \right| \hat{V}_{e-\text{Cs}} \left| n\left(2\frac{1}{2}\right)\frac{5}{2}\frac{1}{2},33 \right\rangle, \tag{S16}$$

which evaluates to

$$V^{3,3}_{5/2,1/2}(R,\theta) = 2\pi\left\{ \left(\frac{21}{80}y_{n0}(R) + \frac{27}{80}y_{n1}(R)\right)|Y_{20}(\theta,0)|^2 + \left(\frac{1}{40}y_{n0}(R) + \frac{3}{8}y_{n1}(R)\right)|Y_{21}(\theta,0)|^2 \right\}. \tag{S17}$$

As we are interested in just the approximate structure of these potential surfaces, we consider now the limit $y_{n0}(R) \to 0$. This is valid as the singlet $S$-wave scattering length in Cs is nearly 20 times smaller than the triplet $S$-wave scattering length. For $\theta=0$, this simplifies to

$$V^{3,3}_{5/2,1/2}(R,0) = 2\pi \cdot \frac{3}{5}\left[\frac{9}{16}y_{n1}(R)\right]\frac{5}{4\pi}.$$

The equivalent expression for the $m_j = 5/2$ potential surface is

$$V^{3,3}_{5/2,5/2}(R,\theta) = 2\pi\left[\frac{9}{16}y_{n1}(R)\right]|Y_{22}(\theta,0)|^2. \tag{S18}$$

In this perturbative limit, these two potential surfaces share the same radial term shown in square brackets.

### Connection to the full basis diagonalization

This suggests that we can use the potential energy curve for the aligned ($\theta=0$) Rydberg $|56D_{5/2}, m_j=1/2, m_f=3\rangle$ state which can be computed with the full basis diagonalization. Let $U_{3,1/2}(R)$ be the potential curve calculated from

6that calculation. It's clear that $U_{3,1/2}(R) \approx V^{3,3}_{5/2,1/2}(R,0)$, where the approximation is only due to the limited basis and lack of spin-orbit scattering included in the calculation of $V^{3,3}_{5/2,1/2}(R,0)$. Thus, within this approximation,

$$\left[\frac{9}{16}y_{n1}(R)\right] = \frac{2}{3}U_{3,1/2}(R), \tag{S19}$$

and hence

$$V^{3,3}_{5/2,1/2}(R,\theta) \to 2\pi\left\{\frac{2}{5}U_{3,1/2}(R)|Y_{20}(\theta,0)|^2 + \frac{2}{3}\cdot\frac{2}{3}U_{3,1/2}(R)|Y_{21}(\theta,0)|^2\right\}$$

$$= 2\pi\frac{2}{5}U_{3,1/2}(R)\left\{|Y_{20}(\theta,0)|^2 + \frac{10}{9}|Y_{21}(\theta,0)|^2\right\} \tag{S20}$$

$$V^{3,3}_{5/2,5/2}(R,\theta) \to 2\pi\frac{2}{3}U_{3,1/2}(R)|Y_{22}(\theta,\phi)|^2. \tag{S21}$$

It is clear from the angular dependence of these surfaces that the rotational and vibrational degrees of freedom no longer separate, as in the Rydberg $n$S molecules. Rather, the Rydberg $n$D molecules will align alongside or perpendicular to the magnetic field axis. The direction of this alignment depends on the functional form of $V_\mu(R,\theta)$. For the Rydberg state $n\text{D}_{5/2}$, $m_j = 1/2$, both configurations are possible due to the existence of both $Y_{20}$ and $Y_{21}$ spherical harmonics in the potential surface decomposition. Only perpendicular alignment is possible in the Rydberg $n\text{D}_{5/2}$, $m_j = 5/2$ case due to the $Y_{22}$ spherical harmonic.

Some discussion is in order regarding the accuracy of these calculations. Fig. S2(a,b) show cuts of the potential surfaces as a function of $\theta$, with $R$ fixed to the minimum of the outermost well. These are computed from the two different calculations discussed here - the dashed black curves show the PEC obtained by diagonalization of the full (reduced subspace) $n$D Rydberg molecule Hamiltonian, while the purple curves show the result of perturbation theory. The dotted gray curves show directly the functional form obtained from the spherical harmonics found in Eqs. S20-S21. Although the overall angular structure is accurately obtained (in the $m_j = 5/2$ case, the molecular states are aligned perpendicularly to the magnetic field axis, while in the $m_j = 1/2$ case the molecular states are predominantly aligned along the magnetic field axis), there are clear differences in the details of the $m_j = 1/2$ case near $\theta = \pi/2$. The overall depths of the surfaces obtained in perturbation theory are around 10-15% deeper than those obtained using the more accurate diagonalization.

The two most important takeaways from this are: (1) that the perturbative potential surfaces fail to properly describe the perpendicularly aligned $m_j = 1/2$ states, although the qualitative description of the molecular states remains quite good, (2) there is already a deviation of about 10% between the diagonalization result and perturbation theory even when the angular structure is preserved. When we use the radial function obtained from the full diagonalization (the $U$ function from above) we see that this gap grows to about 15%. We therefore rescale the potential depth for the $56\text{D}_{5/2}$, $m_j = 5/2$ calculation by 15%. There is no deviation for the $m_j = 1/2$ states, so no correction is applied there.

### Evaluation of binding energies and vibrational wave functions

Since we now have adiabatic potential energy surfaces, $V_\mu(R,\theta) \equiv V^{f,m_f}_{j,m_j}(R,\theta)$, we expand the total rovibrational wave function into spherical harmonics,

$$\psi^{\mu,m_{\tilde{l}}}_v(R,\theta) = \sum_{\tilde{l}=|m_{\tilde{l}}|}^{\infty}\frac{\psi^{\mu,m_{\tilde{l}}}_{v,\tilde{l}}(R)}{R}Y_{\tilde{l}m_{\tilde{l}}}(\theta,0), \tag{S22}$$

where $m_{\tilde{l}}$ is a good quantum number due to the cylindrical symmetry. Inserting this wave function into the nuclear Schrödinger equation and projecting onto spherical harmonics yields the set of coupled equations to solve in order to obtain the molecular wave functions and binding energies,

$$\sum_{\tilde{l}'}\delta_{\tilde{l}'\tilde{l}}\left(-\frac{1}{2\mu_{\text{red}}}\frac{d^2}{dR^2} + \frac{\tilde{l}(\tilde{l}+1)}{2\mu_{\text{red}}R^2} - E_v\right)\psi^{\mu,m_{\tilde{l}}}_{v\tilde{l}}(R)$$
$$= \sum_{\tilde{l}'}\langle\tilde{l}'m_{\tilde{l}}|V_\mu(R,\theta)|\tilde{l}m_{\tilde{l}}\rangle\psi^{\mu,m_{\tilde{l}}}_{v,\tilde{l}}(R). \tag{S23}$$

We solve these equations by expanding the radial wave functions into a B-spline basis. We include angular momenta $|m_{\tilde{l}}| \leq \tilde{l} \leq 50$ to ensure convergence. Some characteristic molecular states obtained from this calculation in the $56\text{D}_{5/2}$, $m_j = 1/2$ and $56\text{D}_{5/2}$, $m_j = 5/2$ potential surfaces are depicted in S2 (c) and S2(d), respectively. The contours of the fixed probability density are plotted in Cartesian coordinates. In the $m_j = 1/2$ case,

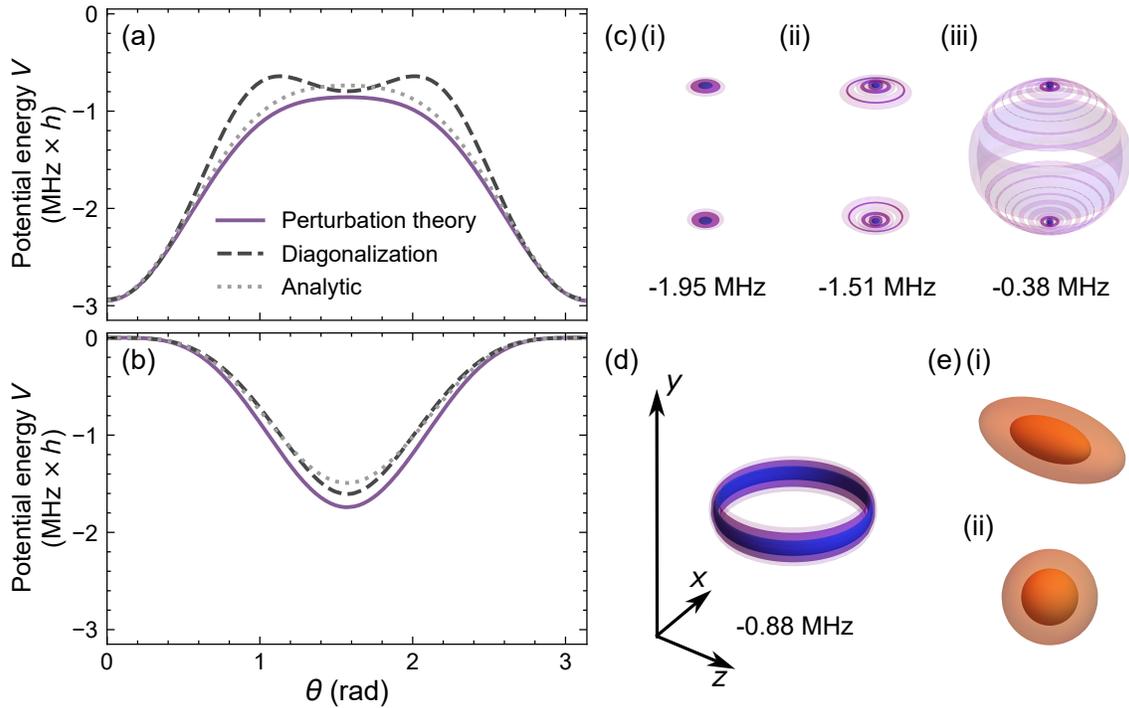

FIG. S2. Qualitative features of Rydberg $n$D state molecules. (a,b) Cuts through the minimum of the radial potential (at $R = 5700\,a_0$) of the two $56D_{5/2}$ potential surfaces considered here, with $m_j = 1/2$ shown in (a) and $m_j = 5/2$ in (b). The dashed black curves show the potential curve resulting from diagonalization of the reduced subspace; the purple curves show the perturbative result including only the diagonal term, and the gray dashed curve shows just the functional form obtained analytically from the spherical harmonics. Panel (c) shows $56D_{5/2}, m_j = 1/2$ molecular states. These are bound in the outermost radial well, and, except for the state in (iii), are aligned tightly to the magnetic field direction. The state in (iii) has also significant probability in the x-z plane (see axes in (d) for orientation), and has a stronger Franck-Condon factor with the more isotropic initial state than either (i) or (ii). (d) shows a $56D_{5/2}, m_j = 5/2$ molecular state, also bound in the outermost radial well but now, due to the different alignment of the potential surface, tightly aligned perpendicular to the magnetic field. (e) shows two different initial states: the anisotropic initial state of the pair of atoms in the same tweezer, and the approximate spherically symmetric state used for the $n$D FCF calculations.

Fig. S2(c), the two more deeply bound states, (i) and (ii), are tightly localized along the magnetic field axis. The state at $-0.38$ MHz leaks out of the deep potential wells at $\theta = 0, \pi$ and has significant amplitude in the x-z plane. The ground states of the $m_j = 5/2$ molecule are strongly localized in the plane, perpendicularly aligned to the magnetic field. Fig. S2(e) shows exemplary initial states of the atom pair: (i) shows the anisotropic initial state due to the different radial and axial trap frequencies in the tweezer, while (ii) shows the spherical approximation used for the $n$D-state Franck-Condon factors (see below).

Figure S3 shows the predicted and measured spectra for both Rydberg $n$D states considered. The measured spectra were obtained with both atoms in the same optical tweezer. The theoretical spectra differ notably from the Rydberg $n$S spectra due to the alignment of the molecules. In the latter case, the number of states in the spectrum was determined by the vibrational splitting alone, as the rotational states were split by only a few kHz and are not resolved. Here, each vibrational state gives rise to a series of "angular" vibrational states with a splitting of up to several hundred kHz. For example, the peaks found at $-3.5$, $-3$ and $-2.2$ MHz are all states without a node in the radial direction but with increasingly many nodes in the $\theta$ direction (see Fig. S2(c, i & ii)). The individual states are not resolved in the experiment, but the overall lineshape is a good match.

In the $m_j = 5/2$ measurement reported in Fig. S3(b), two different series are revealed: one with large FCFs and one with a much smaller FCFs. The states with large FCFs are located at relatively small $R$, bound between the sharp butterfly drop and the increasing outer wall of the potential energy surface as $R$ increases. The binding energies of these states are therefore quite sensitive to the $P$-wave phase shifts, but we expect the same qualitative behavior regardless of the specific phase shifts. The large FCFs of these states compared to those states bound in the outermost potential well (the small peaks in the figure) is due to the fact that the initial pair state of the atoms in the same tweezer is smaller than the Rydberg orbit. When the atoms are prepared in separate tweez-

8

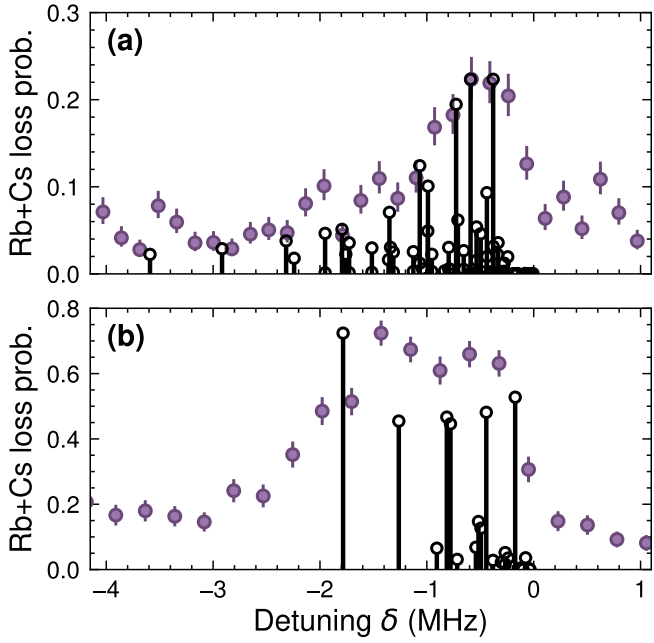

FIG. S3. Comparison of 56D molecular spectra for atoms in the same tweezer. (a) The observed (purple) and predicted spectrum (black) of the $56D_{5/2}, m_j = 1/2$ state. The heights of the black markers indicate the predicted strengths of the molecular transition. The positions of many of the more deeply bound states in the theory calculation are sensitive to the position of the $P$-wave shape resonance, but the overall density of states is not. The high number of states is a direct result of the lack of spherical symmetry of the potential energy surfaces and resulting coupling of rotational levels. (b) The observed (purple) and predicted spectrum (black) of the $56D_{5/2}, m_j = 5/2$ state.

## COUPLING TO MOLECULAR STATES AND LOSS RATES

In this section, we discuss how we calculate the Franck-Condon factors (FCFs). This entails a discussion of the initial pair state of the atoms in the same and separate tweezers, and the evaluation of the overlap integral.

### Calculation of Franck-Condon factors

#### Same Tweezer

When the atoms are trapped within the same tweezer, the initial state of the two ground-state atoms is given to a good approximation by the ground state of a three-dimensional harmonic oscillator with equal radial trap frequencies $\omega_x = \omega_y = \omega_\perp$ and a different axial frequency $\omega_z = \omega_{||}$,

$$\langle \vec{R} | \phi_0 \rangle = \left(\frac{\mu}{\pi}\right)^{3/4} (\omega_\perp^2 \omega_{||})^{1/4} \quad \text{(S24)}$$
$$\times \exp\left(-\frac{\mu}{2}\left[(x^2+y^2)\omega_\perp + z^2 \omega_{||}\right]\right)$$

In spherical coordinates, this has the form

$$\langle \vec{R} | \phi_0 \rangle = A f(R) \exp\left[b(R) \cos 2\theta\right], \quad \text{(S25)}$$

where

$$A = \left(\frac{\mu}{\pi}\right)^{3/4} (\omega_\perp^2 \omega_{||})^{1/4} \quad \text{(S26)}$$

$$f(R) = \exp\left(-\frac{\mu(\omega_\perp + \omega_{||})R^2}{4}\right) \quad \text{(S27)}$$

$$b(R) = -\frac{\mu R^2}{4}(\omega_{||} - \omega_\perp). \quad \text{(S28)}$$

The molecular wave function is given by Eq. S22. Recall that the projection of the molecular rotation onto the magnetic field direction, $m_{\tilde{l}}$, is a good quantum number due to cylindrical symmetry. For the $nS$ Rydberg states, the Born-Oppenheimer potential is considered to be spherically symmetric even in the presence of the magnetic field, and $\tilde{l}$ is also a good quantum number.

The FCF for atoms in the same tweezer is then

$$f_v = \langle \phi_0 | \psi_v^{\mu, m_{\tilde{l}}} \rangle = \sum_{\tilde{l}} \int A f(R) e^{b(R) \cos 2\theta} \frac{\psi_{v,\tilde{l}}^{\mu, m_{\tilde{l}}}(R)}{R} Y_{\tilde{l} m_{\tilde{l}}}(\hat{R}) \mathrm{d}^3 R$$

$$= 2\pi A \delta_{m_{\tilde{l}},0} \sum_{\tilde{l}} \int [R f(R)] \psi_{v,\tilde{l}}^{\mu, m_{\tilde{l}}}(R) \left[\int e^{b(R) \cos 2\theta} Y_{\tilde{l} m_{\tilde{l}}}(\theta, 0) \sin\theta \mathrm{d}\theta\right] \mathrm{d}R$$

Note that we keep the sum over $\tilde{l}$ regardless of the spherical isotropy of the molecular state because the near de-

ers, as in Fig. 4(a,c) in the main text, only these states have non-negligible FCF. This leads to the considerable narrowing seen in the experimental spectra.

generacy of rotational states in the case of Rydberg $n$S states requires us to sum over final states. By symmetry, terms with odd values of $\tilde{l}$ all vanish, and they drop off rapidly in $\tilde{l}$ such that the sum can be truncated by $\tilde{l} < 10$.

### *Separate Tweezers*

We are mostly interested in a qualitative understanding of the FCF for atoms in separate tweezers, and hence make the simplifying assumption that the tweezer potential is spherically symmetric with a trap frequency $\omega$. In this case, we can write the initial state of the two ground-state atoms as the ground-state of a shifted harmonic oscillator,

$$\langle \vec{R} | \phi_0 \rangle = \mathcal{N} \exp[-\alpha|\vec{R} - \vec{R}_0|^2], \tag{S29}$$

where $\alpha = \mu_{\rm red}\omega/2$ and $\mathcal{N} = \sqrt{8\sqrt{\frac{2\alpha^3}{\pi}}}$. This state can be expanded into spherical harmonics, giving a convenient form of this in spherical coordinates

$$\phi_0(\vec{R}) = 4\pi\mathcal{N}e^{-\alpha(R^2+R_0^2)} \sum_{l,m} i_l(2\alpha R R_0) Y_{lm}(\hat{R}) Y_{lm}^*(\hat{R}_0), \tag{S30}$$

where $i_l(x)$ is the modified spherical Bessel function. Since the molecular state is also an expansion into spherical coordinates, the FCF is easily evaluated as

$$\begin{aligned} f_v &= \int \phi_0(\vec{R}) \psi_v^{\mu, m_{\tilde{l}}}(\vec{R}) d^3R \\ &= 4\pi\mathcal{N} \sum_{l=|m_{\tilde{l}}|}^{\infty} Y_{lm_{\tilde{l}}}^*(\hat{R}_0) \\ &\quad \times \int R i_l(2\alpha R R_0) e^{-\alpha(R^2+R_0^2)} \psi_{v,\tilde{l}}^{\mu, m_{\tilde{l}}}(R) dR. \end{aligned} \tag{S31}$$

This last radial integral also decays rapidly as a function of $l$. Unlike the case of atoms in the same trap, this is non-zero for final molecular states with $m_{\tilde{l}} \neq 0$.

For the Rydberg $n$D state calculations when the initial state is a pair of atoms in the same tweezer, we also use this spherical approximation as the initial state is aligned perpendicularly to the magnetic field axis, destroying any cylindrical symmetry (c.f. Fig. S2(e)).

### Modelling the photoassociation rate

To extract the photoassociation rate $\Gamma_{\rm PA}$ we fit the pair loss probability in Fig. 3(b) of the main text with

$$P_{\rm Zero}(t) = A \exp(-\Gamma_{\rm PA} t) + c, \tag{S32}$$

where $t$ is the Rydberg pulse duration and $A$ is the contrast of the loss feature. The contrast is dictated by the state preparation fidelity of the atom pair, the probability to lose the molecule from the tweezers and off-resonant scattering of the Rydberg lasers. The parameter $c$ is an offset accounting for background pair loss in the absence of the Rydberg light, which occurs due to inelastic collisions between Rb and Cs atoms which are not prepared in the lowest energy pair state $|f=1, m_f=1\rangle_{\rm Rb} |f=3, m_f=3\rangle_{\rm Cs}$ in every experimental run.

In the limit of fast molecular decay, $\gamma \gg \Omega_v$, we expect $\Gamma_{\rm PA} \propto \Omega_v^2/\gamma$ [S24, S25], where $\gamma$ is the decay rate of the Rydberg molecule. Therefore, for constant Rydberg laser parameters, we expect $\Gamma_{\rm PA} \propto f_v^2$. The results in Fig. 3(d) and (e) of the main text show qualitatively good agreement with our expectations. However, quantitatively we observe smaller $\Gamma_{\rm PA}$ than expected based on this simple picture. These deviations become large as $\Omega_v$ or the laser linewidth approaches $\gamma$. The intrinsic linewidth of the molecular transition is assumed to be dominated by radiative decay of the Rydberg atom, $\tau_r = 27\,\mu$s (at $T = 300$ K) [S2]. This yields a linewidth of 6 kHz which is smaller than our laser linewidth of 30(10) kHz extracted from the dephasing of Rabi oscillations on the atomic line (see Fig. S4). For an atomic Rabi frequency of $\Omega_{\rm Rb} = 77$ kHz and the maximum predicted FCF of $f_v = 0.06$, we expect $\Omega_v = 4.6$ kHz, which is comparable to $\gamma$. This, combined with the broader laser linewidth leads to an overestimation of $\Gamma_{\rm PA}$ using the approximation described previously.

### EXPERIMENTAL METHODS

### Preparation of atom pairs

We start each experimental run by loading two 1D arrays of optical tweezers which trap Rb and Cs atoms [S26, S27]. We rearrange each of these 1D arrays in order to prepare Rb and Cs atoms in spatially separated optical tweezers at wavelengths 817 nm and 1066 nm respectively. The Rb and Cs atoms are cooled to the motional ground state of their traps with Raman sideband cooling before being prepared in the hyperfine states $|f=1, m_f=1\rangle_{\rm Rb} |f=3, m_f=3\rangle_{\rm Cs}$ [S28]. For excitation to Rydberg $n$D states we prepare the atom pair in the hyperfine states $|f=2, m_f=2\rangle_{\rm Rb} |f=3, m_f=3\rangle_{\rm Cs}$.

Each pair of tweezers containing Rb and Cs is separated from the next pair by at least 1.5× the Rydberg blockade radius. When detailed spectroscopy of the atomic transition is required, we increase the likelihood of preparing only an Rb atom by deliberately placing a Cs atom in every other tweezer.

### Experimental statistics

To obtain statistics, we repeat each experimental sequence many times. Data points in figures show the average probabilities from these repeats and error bars show the $1\sigma$ binomial confidence intervals, calculated using the Jeffreys prior [S29–S31]. For each pair of 817 nm and 1066 nm tweezers in the array, we postselect on the presence of a Rb+Cs atom pair or a Rb atom (only) at the start of the sequence. This postselection allows us to gather statistics for both scenarios—preparing Rb+Cs atoms or only Rb atoms—within a single set of experimental runs using the same sequence. For each experimental datapoint in the main text, we typically repeat the experimental sequence $\sim 150$ times.

### Rydberg excitation scheme

The two-photon Rydberg excitation scheme used in this work is shown in Fig. 1 of the main text and the experimental setup is as described in our previous work [S27]. For excitation to $n$S Rydberg states, the Rb atoms are first prepared in the hyperfine ground state $|g\rangle_{\rm Rb} = |5^2{\rm S}_{1/2}, f=1, m_f=1\rangle_{\rm Rb}$. This state is coupled to $|r\rangle_{\rm Rb}$ via the off-resonant intermediate state $|e\rangle_{\rm Rb} = |6^2{\rm P}_{3/2}\rangle_{\rm Rb}$. The detuning from the intermediate state $\Delta$ is chosen to satisfy $\Delta \gg \Omega_{ge}, \Omega_{er}$, where $\Omega_{ge}$ and $\Omega_{er}$ are the Rabi frequencies for the $|g\rangle_{\rm Rb} \to |e\rangle_{\rm Rb}$ (420 nm) and $|e\rangle_{\rm Rb} \to |r\rangle_{\rm Rb}$ (1020 nm) transitions respectively. When optimizing the Rb*Cs molecular signal, we observed greater contrast in the molecular peaks when using a large intermediate detuning, typically $\Delta = 3.8 - 6.8$ GHz.

The polarization of the 420 nm beam is linear and oriented perpendicular to the quantization axis to drive $\sigma^\pm$ transitions. The 1020 nm beam is circularly polarized and drives $\sigma^+$ transitions. For $n$S Rydberg states, this configuration results in much stronger coupling to states with $m_j = -1/2$ when exciting from the initial state $|5^2{\rm S}_{1/2}, f=1, m_f=1\rangle_{\rm Rb}$. For the Rydberg state $56D_{5/2}$ investigated in Fig. 4 of the main text, we excite the Rb atoms from the spin-stretched state $|5^2{\rm S}_{1/2}, f=2, m_f=2\rangle_{\rm Rb}$ and the strongest coupling is to states with $m_j = 5/2$.

### Measurement of atomic Rabi frequency

In Fig. S4, we show measurements of the atomic Rabi frequency for typical Rydberg laser parameters used in the main text. We plot the probability of Rb loss as a function of the Rydberg pulse duration for light resonant with the state $|36{\rm S}_{1/2}, m_j = -1/2\rangle_{\rm Rb}$. The contrast of the Rabi oscillations is limited by the probability that the Rydberg atom decays before it can be lost from the tweezer. For these relatively short-lived Rydberg states, there is a lower probability of ejection before radiative decay. Only a small fraction of the decayed Rydberg atoms return to the state $|g\rangle_{\rm Rb} = |5^2{\rm S}_{1/2}, f=1, m_f=1\rangle_{\rm Rb}$, leaving a significant portion in other hyperfine states of the electronic ground state. These states are unlikely to couple back to the desired Rydberg state.

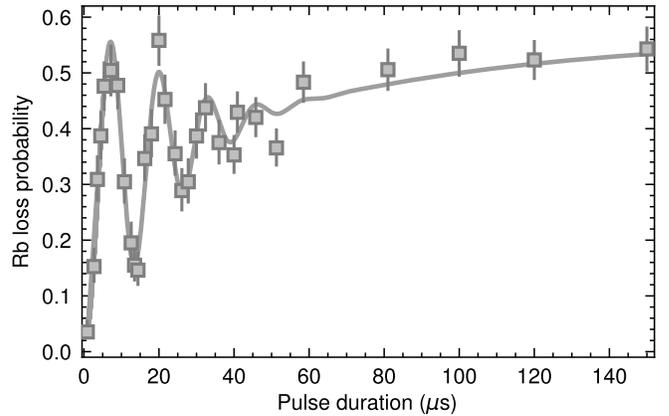

FIG. S4. Measurement of the Rb Rabi frequency for the Rydberg laser parameters used in Fig. 3 of the main text. The probability to lose a Rb atom is plotted as a function of the Rydberg pulse duration. The tweezer intensity was $I = 2$ kW/cm$^2$ and the intermediate state detuning was $\Delta = 6.8$ GHz. The solid line is a fit to the experimental data using a damped sinusoidal model including Rydberg decay.

We fit the data with a damped sinusoidal model which incorporates decay of the Rydberg state. From the fit, we extract a two-photon Rabi frequency $\Omega_{\rm Rb} = 77(1)$ kHz and time constant $38(9)\,\mu$s. This is in good agreement with the expected lifetime of 36S which is $26\,\mu$s ($44\,\mu$s) when including (not including) the effect of blackbody radiation. We also extract the standard deviation of the Rabi frequency to be $7(1)$ kHz from the damping of the Rabi oscillations. Assuming laser frequency noise is the primary cause of this dephasing, we estimate a root-mean-square laser linewidth of $30(10)$ kHz.

### Measured ac Stark shifts

The ac Stark effect causes a shift in the energy levels of atoms or molecules due to their interaction with an oscillating electric field. The energy levels are shifted proportionally to the laser intensity $I$ and the particle's dynamic polarizability $\alpha$, resulting in an energy perturbation $\alpha I$. In the main text, we observe significant ac Stark shifts of the molecular transitions due to the tweezer intensity.

In Fig. S5 we measure the differential stark shift $\Delta \alpha I$ for the atomic transition $|g\rangle_{\rm Rb} \to |r\rangle_{\rm Rb}$ and the molecular transition $|g\rangle_{\rm Rb}|g\rangle_{\rm Cs} \to |36{\rm S}_{1/2}, M_{\rm tot}=5/2, v=0\rangle$. For the atomic transition, the polarizability of both states are

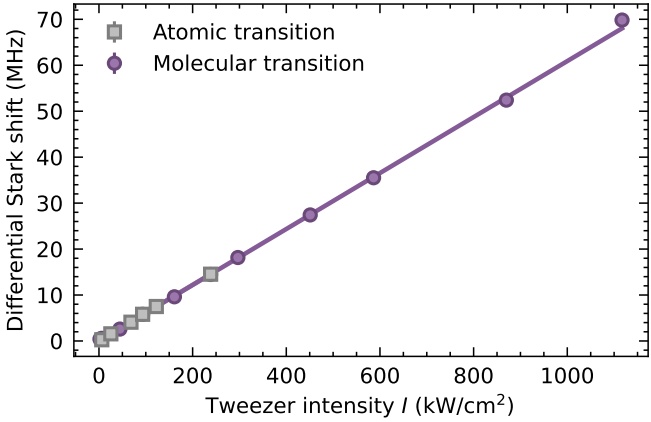

FIG. S5. Differential ac Stark shifts for atomic and molecular transitions in an optical tweezer of wavelength 1065.512 nm. The observed shift in the transition centre relative to the free-space value is shown as a function of tweezer intensity. The grey squares and purple circles represent measurements of the atomic transition $|g\rangle_{\mathrm{Rb}} \to |r\rangle_{\mathrm{Rb}}$ and molecular transition $|g\rangle_{\mathrm{Rb}} + |g\rangle_{\mathrm{Cs}} \to |36\mathrm{S}_{1/2}, M_{\mathrm{tot}} = 5/2, v = 0\rangle$ respectively. The solid lines shown linear fits to the data to extract the gradient.

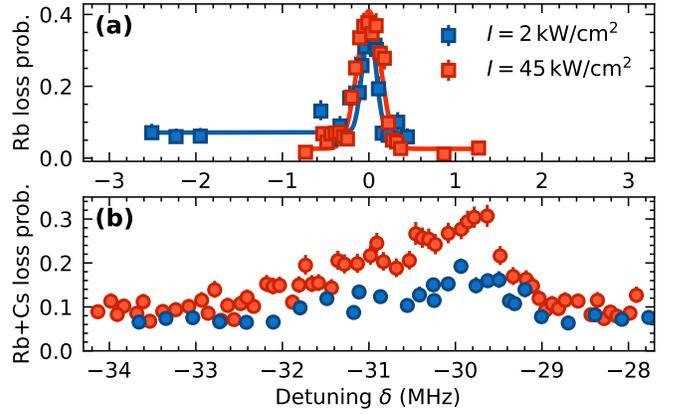

FIG. S6. Comparison of atomic and molecular lineshapes in the vicinity of the state $|36\mathrm{S}, m_j = -1/2\rangle_{\mathrm{Rb}}$. (a) The loss probability of Rb atoms as a function of the detuning from the atomic line. The solid lines show a Gaussian fit to the experimental data. (b) The loss probability of a Rb+Cs atom pair as a function of the detuning from the atomic line. The blue (red) markers show data for a tweezer intensity of 2 kW/cm$^2$ (45 kW/cm$^2$).

known at our tweezer wavelength of 1065.512 nm. The Rb ground-state polarizability is $\alpha_{\mathrm{Rb}} = 685(1) \times 4\pi\varepsilon_0 a_0^3$ [S32] and the polarizability of the Rydberg state, determined by the ponderomotive potential of the Rydberg electron, is $\alpha_r = -550 \times 4\pi\varepsilon_0 a_0^3$ at this wavelength. Using these polarizability values, we predict a gradient of 57.9 kHz/(kWcm$^{-2}$) for the atomic transition. From the measured atomic shift in Fig. S5, we extract a value of 61.1(3) kHz/(kWcm$^{-2}$), where the quoted uncertainty reflects the Type A uncertainty from the linear fit to the data. Comparison of these two values suggests a 6(1)% underestimate of the tweezer intensity from our independent calibration of the tweezer waist and power.

For the molecular transition, we also observe a linear dependence on intensity and extract a gradient of 60.9(3) kHz/(kWcm$^{-2}$) using a linear fit. The values of this fitted gradient and the atomic transition shift are consistent within their uncertainties and we observe no change in the molecule's binding energy as a function of tweezer intensity. This result agrees with the assumption that the molecule polarizability is the sum of its constituent particles' polarizabilities, $\alpha_v = \alpha_{\mathrm{Cs}} + \alpha_r = 613 \times 4\pi\varepsilon_0 a_0^3$. The above measurements suggests that, for this molecular state, the molecule should be trapped in the tweezer.

### BROADENING OF MOLECULAR LINES

In this work, all molecular resonances observed at a magnetic field of 4.78 G in the same optical tweezer exhibited significant asymmetric broadening. Broadening of Rydberg molecule lines has been observed for other Rydberg molecules [S33], where the mechanism behind this broadening was suggested to be residual interactions with other Rydberg atoms or ions in the bulk gas. Using our tweezer setup, we can control precisely the number Rydberg atoms present in our system and we see this broadening persists even when only a single atom pair is prepared in the array.

One potential cause of the broadening is the optical tweezer, which could couple strongly to other molecular states. We investigate the effect of the tweezer broadening on the state $|36\mathrm{S}_{1/2}, M_{\mathrm{tot}} = 5/2, v = 0\rangle$ in Fig. S6. We plot the molecular and atomic line with the same x-axis span to highlight the extent of the broadening. In both cases, we use a shortened Rydberg pulse time to ensure no saturation of the transition. Fitting the atomic line with a Gaussian, we extract a full width at half maximum (FWHM) of 0.18(2) MHz and 0.23(4) MHz for intensities of 2 kW/cm$^2$ and 45 kW/cm$^2$ respectively. These measurements suggest that the tweezer causes little broadening of the atomic line, with the fitted FWHM in good agreement with the expected width based on the Rabi frequency $\Omega_{\mathrm{Rb}} = 77(1)$ kHz and our laser linewidths. However, the molecular peak shows a different behaviour, with a broader, more asymmetric profile. The shape of the molecular line does not appear to be modified by a significant change in the tweezer intensity.

In Fig. S7, we examine two scenarios where the molecular lines are significantly narrower, much closer to the expected widths based on the Rydberg laser linewdiths and other sources of experimental noise. Figure S7(a) compares the molecular state $|48\mathrm{S}_{1/2}, M_{\mathrm{tot}} = 5/2, v = 0\rangle$ at two magnetic fields 4.78(1) G and 181.7(1) G. The $v = 0$

feature at low field shows similar asymmetric broadening to that of $|36S_{1/2}, M_{tot} = 5/2, v = 0\rangle$, while at higher magnetic fields it exhibits much less asymmetry. The FWHM extracted from a Gaussian fit to the spectrum at 181.7 G is 0.5(1) MHz, which is significantly narrower than at 4.78 G, but is still around twice as broad as we would expect for our laser linewidths. A similar behavior is found for all other observed lines.

Figure S7(b) shows spectra for the molecular state $|56D_{5/2}, M_{tot} = 11/2, v = 0\rangle$ (same state as in Fig. 4 of the main text) in the same or separate tweezers at 4.78 G. For the same tweezer case, we observe a similar lineshape to other molecular states at 4.78 G. However, for atoms trapped in separate tweezers, we observe much narrower lines. The fitted FWHM of the narrower peak is 0.2(1) MHz, which is much closer to the broadening expected based on known sources of experimental noise. The same behavior was observed for the state $|58S_{1/2}, M_{tot} = 5/2, v = 0\rangle$ which was the only other state explored in separate tweezers.

The origin of this behavior is currently unknown. The change in structure and width for large magnetic fields suggests some finer structure which is near-degenerate at low magnetic fields and well-resolved at larger magnetic fields. We have compared our measurements to theoretical predictions for molecular lines with different values of $M_{tot}$, corresponding to other $m_f$ states of the Cs atom. The predicted line positions cannot explain the observed lineshapes as they are shifted to the blue. In addition, the very weak mixing of electronic states due to the scattering interaction would only lead to a weak coupling to these states. Another possibility is broadening from the rotational structure of the molecule but this is unlikely due to the small rotational constant of a few kHz.

We have also ruled out stray electric fields as a cause by performing Stark spectroscopy at $n = 83$. The electric field sensitivity of this state is over $300\times$ that of the atomic state 36S. Any broadening from electric field noise is $\ll 1$ MHz at $n = 83$ and the stray electric field is measured to be 0.04(1) V/cm.

One possible cause of the broadening could be from off-resonant couplings of the Rydberg light. This could couple the molecule to a continuum of states above $|e\rangle_{Rb}$ which would give rise to an asymmetric Fano-like profile. A similar effect was recently observed for Rydberg macrodimers in Ref. [S34]. At large magnetic fields the state $|e\rangle_{Rb}$ is in the Paschen-Back regime, leading to a modification of the relevant couplings.

### DISPLACEMENT OF ATOMS IN SEPARATE TWEEZERS

In Fig. 4(c) of the main text, we measure the molecule formation probability as a function of tweezer separation along the x-direction. In this direction, which is perpendicular

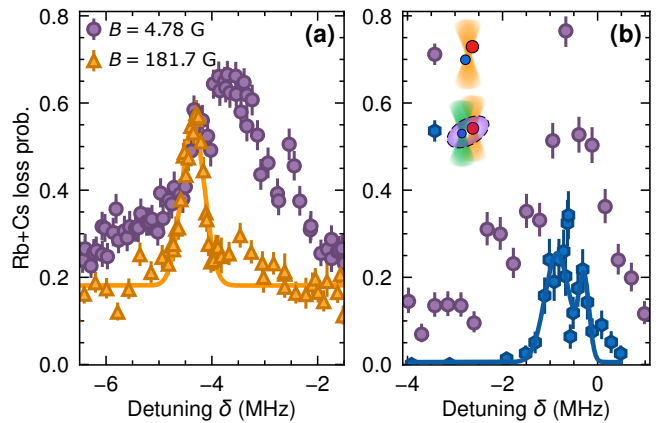

FIG. S7. Suppression of molecular line broadening under specific experimental conditions. In both plots, the pair loss probability is plotted against detuning from the atomic transition. (a) Effect of magnetic field on the spectra of the molecular state $|48S_{1/2}, M_{tot} = 5/2, v = 0\rangle$. Purple circles represent the spectrum at $B = 4.78(1)$ G, while yellow triangles correspond to $B = 181.7(1)$ G. The binding energy of this molecular state is predicted to be $-4$ MHz at both these values of the magnetic field. (b) Effect of confining the Rb+Cs atom pair in the same tweezer (purple circles) compared to separate tweezers (blue hexagons) for the molecular state $|56D_{5/2}, M_{tot} = 11/2, v = 0\rangle$ is shown.

dicular to the direction of tweezer-light propagation, we estimate the separation at sub-micron scales by careful calibration of an acousto-optic deflector (AOD) as described in our previous work [S27]. Briefly, the AOD used to control the position of the 817 nm tweezer, which traps the Rb atom, is calibrated by recording multiple fluorescence images of a Rb atom trapped in the tweezer for a range of radiofrequency drive frequencies. This procedure, along with knowledge of the magnification of our imaging system, allows us to calibrate distance moved in the object plane as a function of the drive frequency of the AOD.

The observed loss maxima in Fig. 4(c) of the main text occur at tweezer separations smaller than the predicted bond length. This discrepancy is most likely due to the non-linear relationship between the *atomic* separation and the *tweezer* separation. At the sub-micron separations used in this work, the optical potential experienced by one atom is significantly perturbed by the other tweezer. The 817 nm tweezer is attractive for Rb but repulsive for Cs. This perturbation results in a complex dependence of atomic positions on the tweezer separation and the intensities of both tweezers. To account for this, we perform numerical simulations of the total potential experienced by each species.

Figure S8 shows the result of these simulations, which use the measured trap waists and Rayleigh lengths of our tweezers. For a given tweezer displacement, the location of the potential minima of each species is determined.

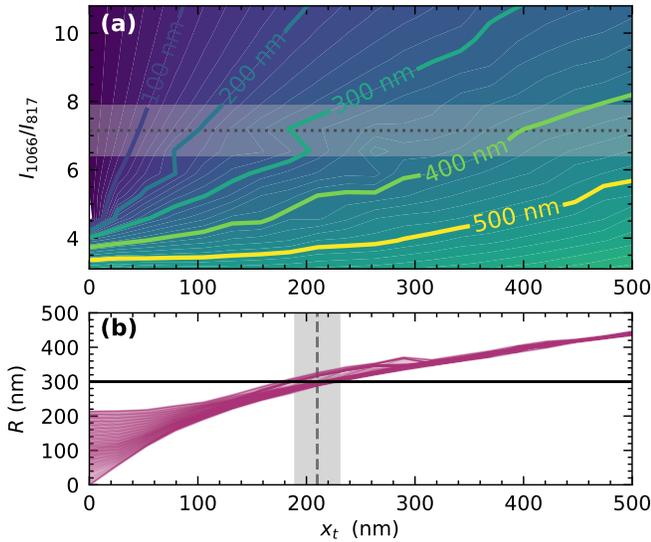

FIG. S8. Simulations of the displacement between Rb and Cs atoms, $R$, for the optical tweezer potentials used in Fig. 4 of the main text. (a) Contour plot of the atomic separation as a function of the intensity ratio of the tweezers $I_{1066}/I_{817}$ and the tweezer displacement along the x-axis $x_t$. The dotted horizontal line indicates the intensity ratio used in the experiment, $I_{1066}/I_{817} = 7.2(7)$, with the shaded region representing its uncertainty. (b) Atomic displacement as a function of tweezer separation for $I_{1066}/I_{817} = 7.2$. The purple lines show simulated atomic separations for various tweezer misalignments along axes perpendicular to the x-axis. The black horizontal line shows the predicted bond length $R_v = 300$ nm for the state $|56D_{5/2}, M_\mathrm{tot} = 11/2, v = 0\rangle$. The vertical grey dashed line indicates the fitted tweezer separation from Fig. 4 of the main text, with the shaded region showing the uncertainty.

The separation between these minima is $R$. Figure S8(a) shows contours of the atomic separation $R$ as a function of the intensity ratio $I_{1066}/I_{817}$ and tweezer separation along the x-axis $x_t$. At smaller intensity ratios, corresponding to larger intensities of the 817 tweezer, the 817 nm tweezer pushes the Cs atom away from the Rb atom, increasing $R$. The dotted line shows the intensity ratio $I_{1066}/I_{817} = 7.2(7)$ used in the experiments in Fig. 4 of the main text. For this value, the atomic separation was around 300 nm for a tweezer displacement of 200 nm.

Another factor effecting the atomic separation is tweezer misalignment along the other axes. Figure S8(b) shows $R$ as a function of $x_t$ for different displacements along the y- and z-axes. Maximum displacements of 50 nm (y) and 100 nm (z), reflecting experimental uncertainty in the overlap of the tweezers, result in a variation of $R$ of approximately 40 nm. However, the dominant factor remains the intensity ratio. Using $I_{1066}/I_{817} = 7.2$, the predicted atomic separation is in good agreement with the predicted bond length $R_v = 300$ nm for a tweezer displacement of $x_t = 210(20)$ nm.


\fi

\end{document}